\documentclass[12pt,preprint]{aastex}

\newcommand{\simgt}{\lower.5ex\hbox{$\; \buildrel > \over \sim \;$}}
\newcommand{\simlt}{\lower.5ex\hbox{$\; \buildrel < \over \sim \;$}}
\newcommand{\tp}{\hspace{-1mm}+\hspace{-1mm}}
\newcommand{\tm}{\hspace{-1mm}-\hspace{-1mm}}
\newcommand{\gIII}{I\hspace{-.3mm}I\hspace{-.3mm}I}

\def\bbeta{\mbox{\boldmath $\beta$}}
\def\btheta{\mbox{\boldmath $\theta$}}
\def\bnabla{\mbox{\boldmath $\nabla$}}
\def\bk{\mbox{\boldmath $k$}}
\newcommand{\cA}{{\cal A}}
\newcommand{\cD}{{\cal D}}
\newcommand{\cF}{{\cal F}}
\newcommand{\cG}{{\cal G}}
\newcommand{\trQ}{{\rm tr}Q}

\newcommand{\paren}[1]{\left( #1 \right)}

\shorttitle{A New Measure for Weak Lensing Flexion}
\shortauthors{Okura, Umetsu, \& Futamase}

\begin{document}


\title{A New Measure for Weak Lensing Flexion}


\author{Yuki Okura\altaffilmark{1}} 
\email{aepgstrx@astr.tohoku.ac.jp}

\author{Keiichi Umetsu\altaffilmark{2}}
\email{keiichi@asiaa.sinica.edu.tw}

\and

\author{Toshifumi Futamase\altaffilmark{1}}
\email{tof@astr.tohoku.ac.jp}


\altaffiltext{1}
 {Astronomical Institute, Tohoku University, Sendai 980-8578, Japan}
\altaffiltext{2}
 {Institute of Astronomy and Astrophysics, Academia Sinica,  P.~O. Box 23-141, Taipei 106,  Taiwan, Republic of China}


\begin{abstract}

We study a possibility to use the octopole moment
of gravitationally lensed images as a direct
measure of the third-order weak gravitational
lensing effect, or the gravitational flexion.
It turns out that there is a natural relation between flexion
and certain combinations of octopole/higher-multipole moments 
which we call 
the Higher Order Lensing Image's Characteristics (HOLICs).
This will allow one to measure directly flexion from 
observable octopole and higher-multipole moments
of background images.
We show 
based on simulated observations
how the use of HOLICs can improve the accuracy
 and resolution of a reconstructed mass map, 
in which we assume Gaussian uncertainties 
in the shape measurements
estimated
using deep $i'$-band data 
of blank fields observed with Suprime-Cam on the Subaru telescope.

\end{abstract}






\section{Introduction}

It is now widely recognized that 
weak gravitational lensing is a unique and valuable tool to 
study the mass distribution of clusters of galaxies 
as well as large scale structure in
the universe 
since it directly measures 
the projected mass distribution of the lens
regardless of the physical state of the system
and the nature of matter content (Bartelmann \& Schneider 2001). 
In the usual treatment of the weak lensing analysis, 
the quadrupole moment of background galaxy images
is used to quantify the image ellipticity.
Then the lensing  properties are extracted from the image ellipticities 
by assuming that source galaxies are randomly oriented in the absence of
gravitational lensing.
In practice we average over a local ensemble of image ellipticities
to estimate the lensing properties.
The local ensemble should contain a sufficient number of 
background galaxies to 
increase the signal-to-noise ratio of local shear measurements,
whereas the region that contains the galaxies should be small enough
to guarantee the constancy of the lensing properties over the region.
The latter condition is necessary in the usual prescription for 
weak lensing 
because it is based on the 
locally linearized lens equation.
On the other hand, the former limits the resolution of 
mass maps reconstructed via weak lensing techniques,
which is of the order 1 arcmin in ground-based observations.

Space-based high-resolution imaging surveys, 
such as the 
Cosmic Evolution Survey (Scoville et al. 2007)
with the {\it Hubble Space Telescope} (HST)
and the proposed {\it Supernova/Acceleration Probe} (SNAP)
wide weak lensing survey (Massey et al. 2004), will provide significant gains 
with a higher surface number density of well-resolved galaxies
due to the small, stable Point Spread Function (PSF),
which will enable high-resolution mapping of the lensing mass
distribution down to an angular resolution of $\sim 0\farcm 1$.
On the other hand, such a small PSF will allow us to resolve 
not only the elliptical component, described by the quadrupole moment,
but also higher-order shape properties of background galaxy images,
which could also 
carry some sort of information of lensing properties.
It may be therefore interesting to see if such higher multipole moments of the
shape are useful for the weak lensing analysis.

There have been some attempts to generalize the weak lensing analysis to
include higher order moments of the light distribution.
Goldberg and Natarajan (2002) 
suggested that higher order effects in gravitational lensing, 
described by the third order derivatives of the lensing potential,
can give rise to octopole moments of the light distribution for background
galaxies.
Goldberg and Bacon (2005)
have further developed their approach and proposed 
a new inversion technique based on the Shapelets formalism 
(Refregier 2003; Refreger \& Bacon 2003;
Massey and Refregier 2005), 
and labeled this third order effect 
as the {\it flexion} of background images.
Irwin and Shmokova (2006) developed a similar analysis method
for measuring the higher order lensing effects and applied this method
to the HST Deep Field North. Recently Irwin, Shmokova, \& Anderson 2006
reported on the detection of lensing signals in the UDF due to small
scale structure using their "cardioid" and "displacement" techniques.
Recently Goldberg \& Leonard (2006)
has extended our HOLICs approach to developed a method
to correct HOLICs for the effect of isotropic Point Spread Function (PSF).

In the present paper, 
following the flexion formalism by Bacon et al. (2006),
we study a possibility to use 
higher multipole moments of 
background source images
for the weak lensing analysis,
and demonstrate via simulations
how such higher order moments can improve the accuracy and resolution
 of a weak lensing mass reconstruction.

The paper is organized as follows.
After briefly summarizing 
the basis of weak lensing and 
the flexion formalism in section 2, 
we introduce higher multipole moments of galaxy images in section 3. 
We define certain combinations of higher multipole moments as HOLICs
(Higher Order Lensing Image's Characteristics) and establish an explicit
relation between flexion and HOLICs. 
In section 4 we present simulations of a weak lensing mass
reconstruction using mock observational data of image ellipticities and HOLICs.
Finally some discussions and comments are given in section 5.

\section{Basis of Weak Lensing and Flexion}

In this section we briefly summarize general aspects of weak lensing and
Bacon et al.'s flexion formalism. 
A general review of weak lensing can be found in
Bartelmann \& Schneider (2001), and we follow the notations and
conventions therein.

\subsection{Local Lens Mapping}

The gravitational deflection of light ray can be described by the lens
equation,
\begin{equation}
\label{eq:lenseq}
\bbeta = \btheta - \bnabla \psi(\btheta),
\end{equation}
where $\psi(\btheta)$ is the effective lensing potential;  $\psi$
is defined 
via the 2D Poisson equation as $\nabla^2\psi(\btheta)
=2\kappa(\btheta)$,
with the lensing convergence $\kappa$.
The convergence $\kappa=\Sigma_m/\Sigma_{\rm crit}$ 
is the dimensionless surface mass density projected on the sky, 
which depends on the lens redshift $z_d$
and the source redshift $z_s$ as well as the background cosmology
through the critical surface mass density
\begin{equation}
\Sigma_{\rm crit}=\frac{c^2}{4\pi G}\frac{D_s}{D_dD_{ds}},
\end{equation}
where $D_d$, $D_s$, and $D_{ds}$ are the angular-diameter distances
from the observer to the deflector, from the observer to the source, and
from the deflector to the source, respectively.
If the angular size of an image is small enough to 
be able to neglect the change of the lensing potential $\psi(\btheta)$,
then we can linearize locally the lens equation (\ref{eq:lenseq}) to have 
$\delta\beta_i={\cal A}_{ij}(\btheta)\delta\theta_j$,
where ${\cal A}_{ij}$ is the
Jacobian matrix of the lens equation,
\begin{equation}
{\cal A}_{ij}\equiv 
\left(\frac{\partial \bbeta}{\partial \btheta}\right)_{ij}
= 
\delta_{ij}-\psi_{,ij} \equiv (1-\kappa)\delta_{ij}-\gamma_{ij}
\end{equation}
where $\gamma_{ij}$ is the trace-free, symmetric shear matrix,
\begin{equation}
\label{eq:shear} 
\gamma_{ij}
:= \left(
\frac{\partial^2}{\partial^i\partial^j}-\frac{1}{2}\nabla^2\delta_{ij}
 \right)
\psi
=\left( 
	\begin{array}{cc} 
	\gamma_1 &  \gamma_2 \\
	\gamma_2 &  -\gamma_1
	\end{array}
\right)
\end{equation}
being defined with the components of gravitational shear
$\gamma=\gamma_1+i\gamma_2$.


\subsection{Gravitational Shear and Quadrupole Shape Moments}

In the usual treatment of weak lensing analysis,
we use quadrupole moments $Q_{ij}$ of the surface brightness distribution
$I(\btheta)$ of
background images for quantifying the shape of the images:
\begin{equation}
\label{eq:Qij}
Q_{ij} \equiv \frac{\int\! d^2\theta\, q_I[I(\btheta)]
\Delta \theta_i \Delta \theta_j}
{\int d^2\theta\,q_I[I(\btheta)]} 
\end{equation}
where  $q_I[I(\btheta)]$ denotes the weight function 
used in the shape
measurement,
and $\Delta \theta_i=\theta_i-\bar{\theta_i}$ is the offset vector
from the image centroid.
Then we define the complex ellipticity $\chi$ as
\begin{equation}
\chi \equiv \frac{Q_{11} - Q_{22} + 2iQ_{12}}{Q_{11} + Q_{22}}. 
\end{equation}
The complex ellipticity $\chi$ transforms under the lens mapping
as
\begin{equation}
\chi^{(s)}=\frac{\chi-2g+g^2\chi^\ast}{1+|g|^2-2Re(g\chi^\ast)}
\label{eq:chis2chi}
\end{equation}
where $g = \gamma/(1 - \kappa)$ is the reduced shear
and $^*$ denotes complex conjugate. 
In the weak lensing limit, 
we neglect 2nd order terms of $g$ and $\chi$,
which yields 
$\chi^{(s)}\approx \chi-2\gamma$.
Assuming the random orientation of the source images, 
we average observed ellipticities over a sufficient number of
images to obtain
\begin{equation}
\langle\chi\rangle\approx 2g \approx 2\gamma.
\end{equation}
The inversion equation from the shear map to the convergence map
is obtained in Fourier space as (Kaiser \& Squires 1993)
\begin{equation}
\label{eq:KS93}
\hat\kappa_{\gamma}(\bk) =
\frac{k_1^2-k_2^2-2ik_1k_2}{k_1^2+k_2^2} \hat\gamma(\bk) \ \ \ \ 
(\bk \neq 0).
\end{equation}

\subsection{Spin Properties}

We define the spin for weak lensing quantities.
A quantity is said to have spin-$N$ if it has the same value after
rotation by $2\pi/N$.
Then, the complex shear $\gamma$, the reduced shear $g$, and the complex
ellipticity $\chi$ are all spin-2 quantities.
The product of spin-$A$ and spin-$B$ quantities has spin-$(A+B)$, and the
product of spin-$A$ and spin-$B^*$ has spin-$(A-B)$.

\subsection{Flexion}

Flexion is introduced to be the third-order lensing effect responsible for 
the weakly skewed and arc-like appearance of lensed galaxies. 
The third-order lensing effect arises from the fact
that the shear and the
convergence are not constant within a source galaxy image.
By taking higher order derivatives of the lensing potential
$\psi(\btheta)$, we can deal with higher order transformations of 
the shape quantities than the complex ellipticity.

Flexion consists of four components of the third-order lensing tensor
${\cal D}_{ijk}={\cal A}_{ij,k}=-\psi_{,ijk}$ 
(see Bacon, Goldberg, Rowe, Taylor 2005).
The first flexion ${\cal F}$ is defined as
\begin{equation}
\label{eq:flexion1}
 {\cal F} = {\cal F}_1 + i{\cal F}_2 
\equiv \partial\partial\partial^*\psi = | {\cal F} | 
e^{i\phi} = \partial \kappa,
\end{equation}
and the second flexion ${\cal G}$ is defined as
\begin{equation}
\label{eq:flexion2}
{\cal G} = {\cal G}_1 + i{\cal G}_2 
\equiv \partial\partial\partial\psi = | {\cal G} | e^{3i\phi} 
= \partial \gamma,
\end{equation} 
where $\partial = \partial_1 + i \partial_2$ is the 
complex gradient operator,
which transforms under rotation as a vector, 
$\partial'=\partial e^{\phi} $, where $\phi$ is 
the angle of rotation.  
Thus ${\cal F}$ has spin-$1$ and ${\cal G}$ has spin-$3$.
The two complex flexion fields satisfy the following consistency relation:
\begin{equation}
\label{eq:consistency}
\partial^* \partial {\cal G} = \partial\partial {\cal F}.
\end{equation}

We then describe the transformation of the shape of a background source 
by expanding the lens equation (\ref{eq:lenseq}) to the second order as
\begin{equation}
d\beta_i \approx {\cal A}_{ij} d\theta_j 
+ \frac{1}{2}{\cal D}_{ijk}d\theta_jd\theta_k.
\end{equation}
The third-order lensing tensor ${\cal D}_{ijk}$ can be expressed 
as the sum of the two terms,  
${\cal D}_{ijk}={\cal F}_{ijk}+{\cal G}_{ijk}$,
with the spin-1 part ${\cal F}_{ijk}$ and the spin-3 part ${\cal G}_{ijk}$:
\begin{eqnarray}
	{\cal F}_{ij1} = -\frac{1}{2}\left(
\begin{array}{@{\,}cc@{\,}}
		3{\cal F}_1  & {\cal F}_2 \\
		 {\cal F}_2  & {\cal F}_1 
	\end{array}
\right) &,& \ \ 
{\cal F}_{ij2} = -\frac{1}{2}\left(
\begin{array}{@{\,}cc@{\,}}
		{\cal F}_2  &  {\cal F}_1 \\
		{\cal F}_1  & 3{\cal F}_2
	\end{array}
\right), \\
{\cal G}_{ij1} = -\frac{1}{2}\left(
\begin{array}{@{\,}cc@{\,}}
		{\cal G}_1 &  {\cal G}_2 \\
		{\cal G}_2 & -{\cal G}_1
	\end{array}
\right)&,& \ \ 
{\cal G}_{ij2} = -\frac{1}{2}\left(
\begin{array}{@{\,}cc@{\,}}
		 {\cal G}_2 & -{\cal G}_1 \\
		-{\cal G}_1 & -{\cal G}_2
	\end{array}
\right).
\end{eqnarray}
Flexion has a dimension of  length$^{-1}$ (or angle$^{-1}$). 
This means that the effect by flexion depends on the source size.
The shape quantities affected by the first flexion ${\cal F}$ alone have
spin-$1$ properties, while those affected by the second flexion ${\cal G}$
alone have spin-$3$ properties.

From equations (\ref{eq:flexion1}) and (\ref{eq:flexion2}),
the inversion equations from flexion to the convergence can be
obtained as follows (Bacon et al. 2006):
\begin{eqnarray}
(\kappa+iB)_{\cal F}&=&\triangle^{-1} \partial^*{\cal F},\\
(\kappa+iB)_{\cal G}&=&\triangle^{-2}
 \partial^*\partial^*\partial^*{\cal G},
\end{eqnarray}
where the complex part $iB$ describes the $B$-mode component that can be
used to test the noise properties of weak lensing data.
An explicit representation for the inversion equations 
is obtained in Fourier space as follows:
\begin{eqnarray}
\label{eq:F2k}
\hat\kappa_{\cal F}(\bk) &=& -i 
\frac{k_1\hat {\cal F}_1 + k_2\hat {\cal F}_2}{k_1^2+k_2^2},
 \\
\label{eq:G2k}	
\hat\kappa_{\cal G}(\bk) &=& 
-i \frac{ 
       \hat {\cal G}_1 (k_1^3 - 3k_1 k_2^2)+
       \hat {\cal G}_2 (3k_1^2 k_2 - k_2^3) 
        }
 {(k_1^2+k_2^2)^2},
\end{eqnarray}
for $\bk\neq 0$.
Further we can combine 
independent mass reconstructions $\hat\kappa_a(\bk)$
$(a=\gamma,{\cal F}, {\cal G})$
linearly in Fourier space
to improve the statistical significance of the $\kappa$ map
with minimum noise
variance weighting:
\begin{equation}
\label{eq:cmap}
\hat\kappa(\bk)=\frac{\sum_a \hat W_{\kappa a}(\bk)\hat{\kappa}_a(\bk)}
{\sum_a \hat W_{\kappa a}(\bk) }, 
\end{equation}
where 
$\hat W_{\kappa a}(\bk) = 1/P^{(N)}_{\kappa a}(\bk)$
with 
noise power spectrum $P^{(N)}_{\kappa a}(\bk)$
of a $\kappa$ map reconstructed
using $a$th observable:
\begin{eqnarray}
P^{(N)}_{\kappa\gamma}(\bk) &=& \frac{P_{\gamma}^{(N)}(\bk)}{2}
= \frac{\sigma^2_{\gamma}}{8\pi n_g}\nonumber\\
\label{eq:power}
P^{(N)}_{\kappa F}(\bk) &=& \frac{P^{(N)}_F(\bk)}{2k^2}=\frac{\sigma^2_{F}}{8\pi n_g \bk^2}\\
P^{(N)}_{\kappa G}(\bk) &=& \frac{P^{(N)}_G(\bk)}{2k^2}=\frac{\sigma^2_{G}}{8\pi n_g \bk^2}\nonumber
\end{eqnarray}
with $P^{(N)}_a(\bk)$ being the shot noise power of $a$th observable,
$\sigma_a$ being the intrinsic dispersion of $a$th observable, and
$n_g$ being the surface number density of background galaxies.
Assuming that errors in $\hat{\kappa}_a(\bk)$ between different
observables are independent, 
the noise power spectrum 
for the estimator (\ref{eq:cmap}) is obtained as
\begin{equation}
P^{(N)}_{\kappa}(\bk)=\frac{1}{\sum_a \hat{W}_a(\bk)}=
\frac{1}{\sum_a 1/P^{(N)}_{\kappa a}(\bk)}.
\end{equation}

\section{Higher multipole moments of images: HOLICs}

In this section we consider higher multipole moments of images and
define useful combinations of them as Higher Order Lensing Image's
Characteristics (HOLICs).
We then derive a simple, explicit relation between flexion and HOLICs.

Higher order moments of images are defined as a straightforward
extension of the quadrupole moment. 
The octopole moment $Q_{ijk}$ and the 16-pole moment $Q_{ijkl}$ are define as follows:
\begin{equation}
\label{eq:Q3}
Q_{ijk} \equiv \frac{\int\!
 d^2\theta\,q_I[I(\btheta)]\Delta\theta_i\Delta\theta_j\Delta\theta_k}{\int\!
 d^2\theta\,q_I[I(\btheta)]},
\end{equation}
\begin{equation}
\label{eq:Q4}
Q_{ijkl} \equiv \frac{\int\! d^2\theta\,q_I[I(\btheta)]\Delta\theta_i
\Delta\theta_j\Delta\theta_k\Delta\theta_l}
{\int\! d^2\theta\,q_I[I(\btheta)]}.
\end{equation}
We first define the normalization factor $\xi$ as
\begin{equation}
\xi \equiv Q_{1111} + 2Q_{1122} + Q_{2222}
\label{eq:xi}
\end{equation}
with spin-0.
Then, we define the following combinations of octopole moments as our HOLICs:
\begin{eqnarray}
\label{eq:zeta}
\zeta &\equiv& \frac{(Q_{111}+Q_{122})+i(Q_{112} + Q_{222})}{\xi},\\
\label{eq:delta}
\delta &\equiv& \frac{(Q_{111}-3Q_{122})+i(3Q_{112} - Q_{222})}{\xi},
\end{eqnarray}
where  
the first HOLICs $\zeta$ has spin-1,
and the second HOLICs $\delta$ has spin-3. 
Note that HOLICs have the dimension of [length]$^{-1}$
(or angle$^{-1}$), the same as flexion does.

Now we are in a position to derive the transformation law of HOLICs
under gravitational lensing.
For this purpose we first derive the relation between 
the source octopole moment $Q^{(s)}_{ijk}$ and 
the image octopole moment $Q_{ijk}$. 
A straightforward calculation leads to  
\begin{eqnarray}
 Q_{ijk}^{(s)}    &\approx&  
 {\cal A}_{il} {\cal A}_{jm} {\cal A}_{kn}Q_{lmn}  
  + \frac{1}{2} 
 \bigl( 
   {\cal A}_{il} {\cal A}_{jm} {\cal D}_{kno} 
+  {\cal A}_{jm} {\cal A}_{kn} {\cal D}_{ilo} \nonumber \\ 
 &&      
+  {\cal A}_{il} {\cal A}_{kn} {\cal D}_{jmo} 
- 4{\cal A}_{il} {\cal A}_{jm} {\cal A}_{kn} F_{o}\bigr) Q_{lmno}
\label{eq:Q_ijk}
\end{eqnarray} 
where we have used the fact that 
the integration measures in the source and image planes 
are related in the following way 
(see Appendix A for detailed calculations):
\begin{eqnarray}
\label{eq:jacob}
d^2\beta 
&=& \left|
     \frac{\partial\bbeta}{\partial\btheta}\right|d^2\theta\nonumber\\
&=& (1-\kappa)^2(1-2F_1 \Delta\theta_1-2F_2 \Delta \theta_2)\,
d^2\theta,
\end{eqnarray} 
to the first order of {\it reduced flexion}
defined as 
\begin{eqnarray}
F&:=&F_1+iF_2=\frac{\cal F}{1-\kappa}\\
G&:=&G_1+iG_2=\frac{\cal G}{1-\kappa}.
\end{eqnarray}
Note that the flexion term from the determinant does not yield a net
contribution to 
the denominators of equations (\ref{eq:Q3}) and (\ref{eq:Q4})
since the coordinate system is taken such that the
first moment of $I(\btheta)$ vanishes:
\begin{equation}
\int\! d^2\beta \, q_I[I(\btheta)] \approx
(1-\kappa)^2 \int\! d^2\theta q_I[I(\btheta)],
\end{equation}
where we have neglected the second order term in $\gamma$.
From this transformation law 
one obtains the desired expressions as
\begin{eqnarray}
\label{eq:zetas2zeta} 
\zeta^{(s)} &=&
\frac{\zeta-2g\zeta^*-g^*\delta -\frac{1}{4}(8F^*\eta
	 +9F+2G\eta^*+G^*\lambda)}
{(1-\kappa)(1-4Re[g^*\lambda]-5Re[F \iota_{I}^*]-Re[G \iota_{\gIII}^*])}, \\
\label{eq:deltas2delta}
\delta^{(s)} &=& 
\frac{\delta-3g\zeta -\frac{1}{4}(10F\eta+7F^*\lambda +3G)}
{(1-\kappa)(1-4Re[g^*\lambda]-5Re[F \iota_{I}^*]-Re[G \iota_{\gIII}^*])},
\end{eqnarray}
where dimensionless quantities $\eta$ and $\lambda$
are defined with 16-pole moments by
\begin{eqnarray}
\label{eq:eta}
\eta &\equiv& 
 \frac{(Q_{1111} - Q_{2222}) +2i(Q_{1112} + Q_{1222})}{\xi},\\ 
\label{eq:lambda}
\lambda &\equiv& 
\frac{(Q_{1111} - 6Q_{1122} + Q_{2222}) +4i(Q_{1112} - Q_{1222})}{\xi},
\end{eqnarray}
with spin-2 and spin-4, respectively;
$\iota_{I}$, $\iota_{\gIII}$ and $\iota_{V}$ are defined with 
$32$-pole moments by
\begin{eqnarray}
\label{eq:iota}
\iota_{I} &\equiv& 
 \frac{(Q_{11111}\tp 2Q_{11122}\tp Q_{12222}) +i(Q_{11112}\tp 2Q_{11222}\tp Q_{22222})}{\xi},\nonumber\\
\iota_{\gIII} &\equiv& 
 \frac{(Q_{11111}\tm 2Q_{11122}\tm 3Q_{12222}) +i(3Q_{11112}\tp
 2Q_{11222}\tm Q_{22222})}{\xi}, \nonumber\\
\iota_{V} &\equiv&
 \frac{(Q_{11111}\tm 10Q_{11122}\tp 5Q_{12222}) +i(5Q_{11112}\tm
 10Q_{11222}\tp Q_{22222})}{\xi},\nonumber\\
{}
\end{eqnarray}
with spin-1, spin-3, and spin-5, respectively.

Assuming that the quantities $g, \eta$ and $\lambda$ are small and 
neglecting higher-order terms containing $32$-pole moments
which are reasonable assumptions on
the weak lensing data, we can approximate the above equations as
\begin{eqnarray}
\label{eq:holic2flex}
\zeta^{(s)}  &\approx&  \zeta-  \frac{9}{4}\frac{\cal F}{1-\kappa},\\
\delta^{(s)} &\approx&  \delta- \frac{3}{4}\frac{\cal G}{1-\kappa}.	
\end{eqnarray}
The formulae (\ref{eq:holic2flex})
make it possible to relate directly
the flexion fields and the HOLICs measurements.
Since the $\zeta$ and $\delta$ are quantities with non-zero spin, namely
quantities with directional dependence,
the expectation values of intrinsic $\zeta$ and $\delta$ are assumed to
vanish, 
\begin{eqnarray}
\label{eq:F2zeta}
\langle \zeta\rangle   &\approx& \frac{9}{4}\frac{\cal F}{1-\kappa} \approx \frac{9}{4} {\cal F},\\ 
\label{eq:G2delta}
\langle \delta \rangle &\approx& \frac{3}{4}\frac{\cal G}{1-\kappa} \approx \frac{3}{4} {\cal G}.
\end{eqnarray} 
Neglecting the flexion term 
in the Jacobian matrix (\ref{eq:jacob}) will lead to a reduction of the 
response $\langle\zeta\rangle/{\cal F}$ 
from $9/4$ to $5/4$ while it will
keep the response $\langle \delta\rangle/{\cal G}$ unchanged,
which was found earlier by Irwin \& Shmakova 2006.
In this way, one can measure directly the flexion fields
${\cal F}(\btheta)$ and ${\cal G}(\btheta)$ 
from the observable HOLICs.
Once we obtain the flexion fields, we can make use of equations
(\ref{eq:F2k}) and (\ref{eq:G2k})
to invert them to the surface mass distribution. 

It is important to note the above relation (\ref{eq:F2zeta}) 
is modified if we take into account 
the fact that the "apparent" center of an image defined by the first
moment  of the image is different from the "actual" center 
mapped by the lens equation from the center of the source. 
We discuss in detail this shift of the centroid in Appendix B.

\section{Simulated Observations}

In order to test the performance of 
mass reconstructions based on HOLICs measurements, 
we generate simulated observations of the 
weak lensing effects, namely $(\chi, \zeta, \delta)$,
that include observational errors as Gaussian uncertainties.
The flexion fields (${\cal F}(\btheta)$, ${\cal G}(\btheta$))
can be used to reconstruct mass maps directly, independent of
information on the shear field $\gamma(\btheta)$ (see \S 2.4).
The equation (\ref{eq:holic2flex}) 
defines the direct, unbiased estimators for the flexion fields,
where the precision of this measurement depends on 
the intrinsic values of HOLICs convolved with the measurement noise.
In the present simulation we do not take into account explicitly the centroid shift
(see Appendix B) but use directly equations (\ref{eq:zetas2zeta})  and
(\ref{eq:deltas2delta}) to calculate the lensed HOLICs 
from the intrinsic shape quantities and lens properties,
which does not require the measurement and removal of the apparent
centroid of an image.

To determine the widths of random Gaussian distributions
for the noise component of HOLICs $(\zeta,\delta)$,
arising from the intrinsic scatter in unlensed HOLICs and
observational noise,
we refer to variances of HOLICs obtained from our preliminary study
of  deep $i'$-band data of $\sim 2\times 4$ ${\rm deg}^2$ 
blank fields
observed with Suprime-Cam on the Subaru telescope 
(T. Yamada, private communication). Each $4 {\rm deg}^2$ data set
consists of $18$ Suprime-Cam pointings with $0\farcs 202$ pixel$^{-1}$.
We used our weak lensing analysis pipeline
based on IMCAT (Kaiser, Squires, \& Broadhurst 1995) 
extended to include
the higher multipole moments in the shape measurements.
We selected 
a sample of $493934$ background galaxies 
with $20 \le i'_{\rm AB} \le 24.5$ mag
in the blank fields, 
corresponding to a mean surface number density of $\bar{n}_g\approx 17\,
{\rm arcmin}^{-2}$.
Here we discarded
stars and all objects for which reliable shape measurements cannot be
found.
In particular, we excluded those small objects whose
half-light radius $(r_h)$ is smaller than $0\farcs 50$
(2.5 pixels).
We note the median value
of stellar $r_h$ over the entire field is 
$\langle r_h^* \rangle_{\rm med}\approx 0\farcs 40$ 
with a dispersion of $0\farcs 08$ 
using $N_*=27958$ stars ($n_{*} \approx 1 \, {\rm arcmin}^{-2}$).
This lower cut-off in the galaxy size is essential 
for us to be able to make reliable 
shape measurements because 
(1) the smaller the object, the noisier its shape measurement
due to pixelization (i.e. discretization) noise, 
in particular for the case of 
higher-order shape moments,
and
(2) the shape of an image whose intrinsic   
size is smaller than or comparable to the size of PSF can be 
highly distorted and smeared.
For example, if the spatial 
distributions of both the source and the PSF are described by
a two-dimensional Gaussian, with half-light radii of $r_{h,0}$ and
$r_h^*$, respectively, then the half-light radius of the PSF-convolved
image is $r_h =\sqrt{r_{h,0}^2+r_h^{*2}}$.
When both the sizes are equal, then $r_h = \sqrt{2} r_h^*$,
which is close to our choice for the lower cut-off in $r_h$.

We estimated the unlensed dispersions of HOLICs to be
$\sigma_{\zeta} \equiv \sqrt{\langle|\zeta|^2 \rangle}
\approx 0.0215 \, {\rm arcsec}^{-1}$
and
$\sigma_{\delta} \equiv 
\sqrt{\langle|\delta|^2 \rangle} \simeq 0.0248 \, {\rm arcsec}^{-1}$,
where noisy outliers responsible for the non-Gaussian tail were
removed in the variance estimation.
After clipping rejections
the number of galaxies in this ``clean sample''
is about $N_{\rm clean}\approx 3.1\times 10^5$
($\bar{n}_g\approx 11 \, {\rm arcmin}^{-2}$),  
and the median value of $r_h$ is 
$\langle r_h^*\rangle \approx 0\farcs 639$.
Using the clean sample we also measured dispersions
of dimensionless HOLICs, $a\zeta$ and $a\delta$,
with $a$ being the characteristic scale of
the observed galaxy (Goldberg \& Bacon 2005; Goldberg \& Leonard 2006). 
We take $a$ to be the half-light diameter,
$a=2r_h$, while  Goldberg \& Bacon (2005) and Goldberg \& Leonard (2006)
chose $a$ to be the semimajor axis for measuring the intrinsic flexion
of galaxies.
We found $\sigma_{a\zeta}=0.0286$ and $\sigma_{a\delta}=0.0344$
with the median size of $\langle a\rangle_{\rm med}=1\farcs 27$.
We note that there is a good agreement between $\sigma_{a\zeta}/\langle
a\rangle_{\rm med}$
($\sigma_{a\delta}/\langle a\rangle_{\rm med}$) and $\sigma_{\zeta}$
($\sigma_{\delta}$), indicating that the measured dispersions are 
effectively weighted by moderately large galaxies with $a\approx 1'$. 
The first HOLICs $\zeta$, which is a spin-1 quantity, is sensitive to
the determination of the centroid, or the first moment, with spin-1.
Thus the error in the centroid determination affects seriously the
estimation of the first HOLICs $\zeta$, while this effect is of second
order for other shape quantities with non spin-1 quantities.
Note that
even though small objects were excluded from the analysis,
no anisotropic/isotropic PSF corrections were applied 
in measuring the HOLICs for the present study, which implies that the
dispersions of HOLICs were underestimated to some degree
due to the isotropic smearing
effect by the atmosphere and the Gaussian weighting in $q[I(\btheta)]$
(see Fig.~2 of Irwin et al. 2006).


In the present study, 
we simply adopt 
for illustration purposes
random Gaussian distributions with
dispersions of 
$\sigma_{\zeta}^{(s)}=\sigma_{\delta}^{(s)}=0.02 \,{\rm arcsec}^{-1}$  
in generating the intrinsic values of HOLICs,
corresponding to flexion dispersions of
$\sigma_{F}^{(s)}\simeq 0.009 \, {\rm arcsec}^{-1}$ and 
$\sigma_{G}^{(s)}\simeq 0.027 \, {\rm arcsec}^{-1}$.
We note that
Bacon et al. (2006) adopted
$\sigma_{F}^{(s)}=\sigma_{G}^{(s)}=0.04 \, {\rm arcsec}^{-1}$
in their simulation.
For the intrinsic
dispersion of ellipticities, we adopt $\sigma_{\chi}^{(s)}=0.4$.
As a lens model we use a mock cluster located at $z_d=0.1$,
which is shown as the dimensionless surface mass density $\kappa$
in the top-left panel of Fig.~1.
The cluster consists of a main halo and five sub-halos
which are described by NFW density profiles with different masses and
concentration parameters.
The field is square-shaped with a side length of $6.4$ arcmin.
We use $N_s=4,096$ sources
($n_g=100 \, {\rm arcmin}^{-2}$)
randomly distributed over the field, with intrinsic shape quantities
$(\chi^{(s)}, \zeta^{(s)}, \delta^{(s)})$
drawn from random Gaussian distributions with dispersion
$(\sigma_{\chi}^{(s)}, \sigma_{\zeta}^{(s)},
\sigma_{\delta}^{(s)})$, respectively. 
We assume the background sources are located at
a single redshift of $z_s=1.2$, which is a fair approximation for the
lens located at a low redshift of $z_d=0.1$.
These observation parameters are
appropriate for a future space-based survey such as the planned weak
lensing survey with the SNAP satellite (Massay et al. 2004),
except that the intrinsic flexion dispersions 
$(\sigma_{F}^{(s)}, \sigma_{G}^{(s)})$
are derived from ground-based weak lensing data with galaxies of
$a =2r_h \approx 1''$.
Hence the values of $(\sigma_{F}^{(s)}, \sigma_{G}^{(s)})$ adopted in
the present study will probably be very optimistic for space-based data
with distant galaxies of $a \sim 0\farcs 5$ (see Fig.~2 of Massey et
al. 2004).\footnote{For a galaxy with a Gaussian profile, 
$a=2r_h = \sqrt{8\ln{2}}\sigma \approx 2.354 \sigma$ with $\sigma$ being
the Gaussian dispersion. Hence, $a\sim  0\farcs 5$ for
$\sigma \sim 0\farcs 2$.} Such small, distant galaxies will have
intrinsic flexion dispersions greater by a factor of a few.
Finally,
we use equations (\ref{eq:chis2chi}), (\ref{eq:zetas2zeta}), and 
(\ref{eq:deltas2delta}) to generate lensed quantities
$(\chi, \zeta, \delta)$ for all sources.

Figure 1 shows reconstructed mass maps of the model cluster using
$(\chi, \zeta, \delta)$ along with the input mass model.
We used the linear inversion equations 
(\ref{eq:KS93}), (\ref{eq:F2k}),  and (\ref{eq:G2k}).
The reconstructed $\kappa$-maps were smoothed 
with a Gaussian filter, where the Gaussian FWHMs
are taken to be $(0\farcm 333, 0\farcm 083,  0\farcm 249)$
for reconstructions using 
$(\gamma, F, G)$, respectively.
Then, the dispersions in the Gaussian-smoothed $B$-mode maps
are obtained as $(\sigma_B(\gamma), \sigma_B(F), \sigma_B(G))
=(0.0284, 0.0236, 0.0570)$.
The $\kappa$ maps reconstructed using HOLICs recover 
substructures better than large-scale structures,
allowing a high-resolution mass reconstruction.
The superior sensitivity of flexion to small scale structure
comes from the $1/k^2$-dependence of the noise power spectrum
in a $\kappa$ reconstruction (see equation [\ref{eq:power}]).
Further, since we have assumed
$\sigma^{(s)}_{\zeta}=\sigma^{(s)}_{\delta}$
whereas $\zeta$ has a three-times larger response to flexion than
$\delta$, the ${\cal F}$-based reconstruction has a three-times better
sensitivity than the $\cal G$-based reconstruction.

In Fig.~2 we show the $\kappa$ map obtained by combining
shear- and flexion-based reconstructions 
using equation (\ref{eq:cmap}). 
On small angular scales the signal is dominated by
flexion,
while it is dominated by gravitational shear on large angular scales.
Since $\hat{W}_{\kappa F}(\bk) > \hat{W}_{\kappa G}(\bk)$ 
for all wavenumbers,
the $\kappa$ map is less weighted by the second flexion for 
the observation parameters adopted in this study
(i.e., $\sigma_{\zeta}=\sigma_{\delta}$).
The rms noise level in the $\kappa$-map from flexion and shear data
is reduced down to 
\begin{equation}
\sigma_B \approx \frac{1}{\sqrt{\sum_{j=\gamma,F,G} 1/\sigma^2_B(j)}}
=0.0173.
\end{equation}

\section{Discussion and Conclusions}

In the present paper, we have studied the possibility to improve 
the weak lensing analysis by utilizing the octopole and higher-multipole
moments of lensed images that carries the third-order weak lensing effect.
By defining proper combinations of octopole moments as 
HOLICs, we have derived explicit relations between 
the flexion fields $(F,G)$
and observable HOLICs $(\zeta,\delta)$. 
In the weak lensing limit, the first flexion $F$ excites 
in lensed images
the first
HOLICs $\zeta$ with spin-1, while the second flexion $G$ excites the
second HOLICs $\delta$ with spin-3.
One can employ the assumption of
random orientation for intrinsic HOLICs of background sources
to obtain an unbiased, direct estimator for flexion,
in a similar manner to the usual prescription for weak lensing.

We have also shown by using simulated observations
how the use of HOLICs can improve the
accuracy and resolution of a reconstructed mass map,
in which we assumed Gaussian uncertainties in the shape measurements
estimated using deep $i'$-band data of blank fields
observed with Subaru/Suprime-Cam.
The gravitational shear and flexion have
different scale-dependence in mass reconstruction errors.
The mass maps reconstructed using HOLICs
recover substructures better than the shear-based reconstruction,
allowing a high-resolution mass reconstruction.
It is shown that an optimal linear combination of individual
mass reconstructions can be formed 
using the statistical weight in Fourier space,
which can improve the statistical significance of weak lensing mass
reconstructions.
In actual observations, on the other hand,
we must apply various shape corrections 
(e.g., isotropic/anisotropic PSF corrections)
to the higher order shape quantities
in order to measure flexion to high precision.  
In particular, the first HOLICs $\zeta$ with spin-1 is highly sensitive to 
the choice of the center-of-image.
These issues will be discussed further in the forthcoming
publications.


\acknowledgments

We are grateful to Yasunori Sato and Toru Yamada for
their providing Subaru blank field data.
We acknowledge useful discussions with
Masahiro Takada and Patrick Koch.
We thank the Suprime-Cam team for their support during
the observation. 
We thank the annonymous referee for invaluable comments and suggestions
which improved the paper significantly,
This work is partially supported by the COE program at Tohoku University.

\appendix
\section*{Appendix A: Calculation of the Jacobian}

We present detailed calculations of the Jacobian (29) in this appendix.
Using the lens equation the Jacobian can be calculated as follows:
\begin{eqnarray}
\label{eq:jacob_app}
d^2\beta &=&
{\rm det}
\left(
\begin{array}{cc}
\partial\beta_1/\partial\theta_1 & \partial\beta_1/\partial\theta_2\\
\partial\beta_2/\partial\theta_1 & \partial\beta_2/\partial\theta_2\\
\end{array} 
\right)d^2\theta\nonumber\\
&=& 
\Biggl(
 \left(
 \cA_{11} + \cD_{111}d\theta_1 + \cD_{112} d\theta_2
 \right)
 \left(
 \cA_{22} + \cD_{122}d\theta_1 + \cD_{222} d\theta_2
 \right)
-\left(
\cA_{12} + \cD_{112}d\theta_1 + \cD_{122} d\theta_2
\right)^2
\Biggr)  d^2\theta\nonumber\\
&\approx&
(1-\kappa)^2 \times \nonumber\\
& & \left(
1 - g - \frac{1}{2}(3F_1 + G_1)d\theta_1 - \frac{1}{2}(F_2 +
G_2)d\theta_2
\right)
\left(
1+g -\frac{1}{2} (F_1 - G_1)d\theta_1
-\frac{1}{2}(3F_2 - G_2) d\theta_2
\right)  d^2\theta\nonumber\\
&\approx&(1-\kappa)^2\paren{1- 2F_1d\theta_1 - 2F_2d\theta_2}d^2\theta.
\end{eqnarray}

\section*{Appendix B: Effect of the Centroid Shift}

In a weak lensing analysis
we quantify the shape of an image by measuring various moments of
the surface brightness distribution $I(\btheta)$, in which 
the moments are calculated with respect to the centroid of the image,
or the center of light, defined by the first moments of $I(\btheta)$:
%
\begin{equation}
\label{eq:centroid}
\bar{\theta}_i \equiv \frac{\int\!d^2\theta\,\theta_i I(\btheta)}{\int\!d^2\theta\, I(\btheta)}.
\end{equation}
However, in general,
this apparent center
can be different from the ``point'' that is mapped 
using the lens equation
from the center of unlensed light.
We refer to this point as the ``true'' center of the image.
The difference between these two centers, namely
the apparent and the true centers,
cause a significant effect in evaluating the first flexion as pointed
out by Goldberg \& Bacon (2005).
For the second flexion the effect of the centroid shift is second order
so that we will ignore it in the present paper.


Let us define the center of the source in the absence of gravitational
lensing as $\bar{\beta}_i$ and the apparent center of the
lensed image as $\bar{\theta}_i$.
We also define the true center of the image that is
mapped from 
$\bar{\beta}_i$ using the lens equation,
as $\theta_i(\bar{\bbeta})$.
Similarly, we define the point in the source plane
that is inversely mapped by the lens equation from $\bar{\theta}_i$ as 
$\beta_i(\bar{\btheta})$.
Then, we have
\begin{eqnarray}
\label{eq:dbeta}
\beta_i(\bar{\btheta})
 - 
\bar{\beta}_i &=&
\bar{\theta}_i - 
\alpha_i(\bar{\btheta}) 
- 
\bar{\beta}_i
=  \bar{\theta}_i - \alpha_i(\bar{\btheta}) 
- 
\frac{\int\! d^2\beta\, \beta_i I^s(\bbeta)}
     {\int\! d^2\beta\, I^s(\bbeta)} \nonumber\\
&=& \bar{\theta}_i - \alpha_i(\bar{\btheta}) - 
\frac{
     \int \! d^2\theta\, (1-2F_j d\theta_j)
     \paren{\theta_i - \alpha_i(\btheta)}I(\btheta)}
    {
     \int \! d^2\theta\, (1-2F_jd\theta_j)I(\btheta)
}
%
\end{eqnarray}
where 
we have used the expression (\ref{eq:jacob_app}) for the Jacobian
and used the relation between the lensed and unlensed surface brightness
distributions, $I(\btheta)=I^s(\bbeta)$, 
with $\beta_i=\theta_i-\alpha_i(\btheta_i)$.
We then expand the deflection angle $\alpha_i$ in equation
(\ref{eq:dbeta})
as 
\begin{equation}
\alpha_i(\btheta) \approx 
 \alpha_i(\bar{\btheta}) + 
 \alpha_{i,j} d\theta_j + \frac{1}{2}\alpha_{i,jk} d\theta_j d\theta_k = 
\alpha_i(\bar{\btheta}) - \paren{\cA - {\bf 1}}_{ij}d\theta_j
- \frac{1}{2}\cD_{ijk} d\theta_jd\theta_k
\end{equation}
with $\theta_i=\bar\theta_i + d\theta_i$. 
Thus, in the first order of the gravitational shear and flexion,
we have 
\begin{eqnarray}
\label{eq:SC}
\beta_i(\bar{\btheta}) - \bar{\beta}_i 
&=& 
  \bar \theta_i - \alpha_i(\bar{\btheta}) \nonumber\\ 
 &&- \frac{\int\! d^2\theta\, (1-2F_ld\theta_l)
         \paren{\bar \theta_i + d\theta_i - \alpha_i(\bar{\btheta}) 
         + \paren{\cA - {\bf 1}}_{ij} d\theta_j 
         + \frac{1}{2} \cD_{ijk} d\theta_j d\theta_k}I(\btheta)
      }
    {
      \int \! d^2\theta\, (1-2F_ld\theta_l)I(\btheta)} \nonumber\\
&=&
  \frac{
    \int\! d^2\theta\, (1-\kappa) 2F_ld\theta_l d\theta_i I(\btheta)}
      {
    \int\! d^2\theta\, I(\btheta)
      } 
 - \frac{1}{2}
   \frac{
      \int\! d^2\theta\, \cD_{ijk}d\theta_j d\theta_k I(\btheta)}
        {
      \int\! d^2\theta\, I(\btheta)} \nonumber\\
&=& \trQ\paren{\frac{3}{2}\cF_i + \frac{5}{4}[\cF^*\chi]_i +
\frac{1}{4}[\cG\chi^*]_i},
\end{eqnarray}
where $\trQ = Q_{11}+Q_{22}$ is the trace of $Q_{ij}$ defined by 
equation (\ref{eq:Qij}).
Thus the displacement from the true to the apparent center
is
$\trQ\paren{\frac{3}{2}\cF_i + \frac{5}{4}[\cF^*\chi]_i +
\frac{1}{4}[\cG\chi^*]_i}$ in the ``source plane'',
and all of the shape quantities in equation (\ref{eq:SC}) 
are measured using the apparent center $\bar{\btheta}$. 
In the zeroth order of the shear and flexion, 
this displacement is magnified by 
$1/(1-\kappa)$ in the image plane, so that we have
\begin{eqnarray}
\theta_i(\bar{\bbeta}) \approx 
 \bar{\theta}_i
 - \trQ \paren{\frac{3}{2} F_i 
 + \frac{5}{4}[F^*\chi]_i + \frac{1}{4}[G\chi^*]_i}
\end{eqnarray}
where $F$ and $G$ are the first and the second flexion, $F={\cal
F}/(1-\kappa)$ and $G/(1-\kappa)$.

Next,  we evaluate the effect of the centroid shift
for measuring the first HOLICs, $\zeta$.
We define the apparent octopole moments (with respect to the apparent
center
$\bar{\btheta}$)
as
\begin{eqnarray}
Q_{ijk} = 
\frac
{\int\! d^2\theta\, 
  d\theta_i d\theta_j d\theta_k I(\btheta)}
{\int\! d^2\theta\, I(\btheta)}
\end{eqnarray}
and the ``true'' octopole moments
(with respect to the true center $\btheta(\bar{\beta})$)
as
\begin{eqnarray}
Q^t_{ijk} &=& 
\frac{
  \int\! d^2\theta\, \left(\theta_i-\theta_i(\bar{\bbeta})\right)
                     \left(\theta_j-\theta_j(\bar{\bbeta})\right)
                     \left(\theta_k-\theta_k(\bar{\bbeta})\right)
        I(\btheta)}
{\int\! d^2\theta\, I(\btheta)}\\
			&=& 
\frac{
  \int\! d^2\theta\, (d\theta_i+\Delta^{\theta}_{i})
                     (d\theta_j+\Delta^{\theta}_{j})
                     (d\theta_k+\Delta^{\theta}_{k})
  I(\btheta)}
 {\int\! d^2\theta\, I(\btheta)}
\end{eqnarray}
where 
$\Delta^{\theta}_{i} = \trQ\paren{\frac{3}{2}F_i + \frac{5}{4}[F^*\chi]_i 
+ \frac{1}{4}[G\chi^{*}]_i}$.
Using the expressions for the moments above,
it is straightforward to 
calculate the corresponding HOLICs 
using the moments with respect to the true center,
and then to have the relation between 
$\zeta^t$ constructed using the true center
and  
$\zeta$ constructed using the apparent center:
\begin{equation}
\label{eq:zeta_true}
\zeta^t = \zeta  + 2\frac{\trQ}{\xi} \Delta^{\theta} 
+ \frac{\trQ}{\xi}[\chi\Delta^{\theta *}]
\approx 
\zeta + \frac{\paren{\trQ}^2}{\xi}\left( 3F + 4F^*\chi 
+ \frac{1}{2}G\chi^{*}\right)
\end{equation}
as well as the relation between $\delta^t$ and $\delta$: 
\begin{equation}
\label{eq:delta_true}
\delta^t =
\delta + 3\frac{\trQ}{\xi}[\chi\Delta^{\theta}]\approx \delta + \frac{9}{2}\frac{\paren{\trQ}^2}{\xi}\chi F,
\end{equation}
up to the the second order of the shear, flexion and 
non 0-spin shape quantities. 
Equations (\ref{eq:zeta_true}) and (\ref{eq:delta_true})
allow us to establish the required relationships 
between the unlensed HOLICs $(\zeta^{(s)}, \delta^{(s)})$
and the lensed HOLICs $(\zeta, \delta)$
(note that equations [\ref{eq:zetas2zeta}] and [\ref{eq:deltas2delta}]
are the relations 
between $\zeta^{(s)}$ and $\zeta^t$,
and between $\delta^{(s)}$ and $\delta^t$, respectively, 
in the notation of this appendix),
\begin{eqnarray}
\zeta^{(s)} &=&
\frac{\zeta-2g\zeta^*-g^*\delta 
-
\frac{1}{4}(8F^*\eta - 16\frac{\paren{\trQ}^2}{\xi}F^*\chi
	 +9F 
   - 12\frac{\paren{\trQ}^2}{\xi}F +2G\eta^* - 2\frac{\paren{\trQ}^2}
    {\xi}G\chi^*+G^*\lambda)}
  {(1-\kappa)(1-4Re[g^*\lambda]-5Re[F \iota_{I}^*]-Re[G
  \iota_{\gIII}^*])}, \\
\delta^{(s)} &=& 
\frac{\delta-3g\zeta -\frac{1}{4}
  (10F\eta - 18\frac{\paren{\trQ}^2}{\xi}F\chi+7F^*\lambda +3G)}
  {(1-\kappa)(1-4Re[g^*\lambda]-5Re[F \iota_{I}^*]-Re[G \iota_{\gIII}^*])}.
\end{eqnarray}
Finally, 
in the first order,
the first flexion is expressed using the observable shape
quantities as
\begin{eqnarray}
\label{eq:F2zeta_cor}
\langle \zeta\rangle
\approx 
\left(
 \frac{9}{4} -3  \left< 
  \frac{(\trQ)^2}{\xi} 
  \right>
\right)\,F.
\end{eqnarray}
One can see the asymptotic behavior of the correction term
in equation (\ref{eq:F2zeta_cor}),
when the lensed image is close to a circle.
For an image with the circular top-hat
brightness profile,
we have $(\trQ/\xi)^2 = 3/4$, so that 
one will measure 
$\zeta=0$ regardless
of the value of $F$.

\clearpage



\begin{figure}
\epsscale{1.0}
\plotone{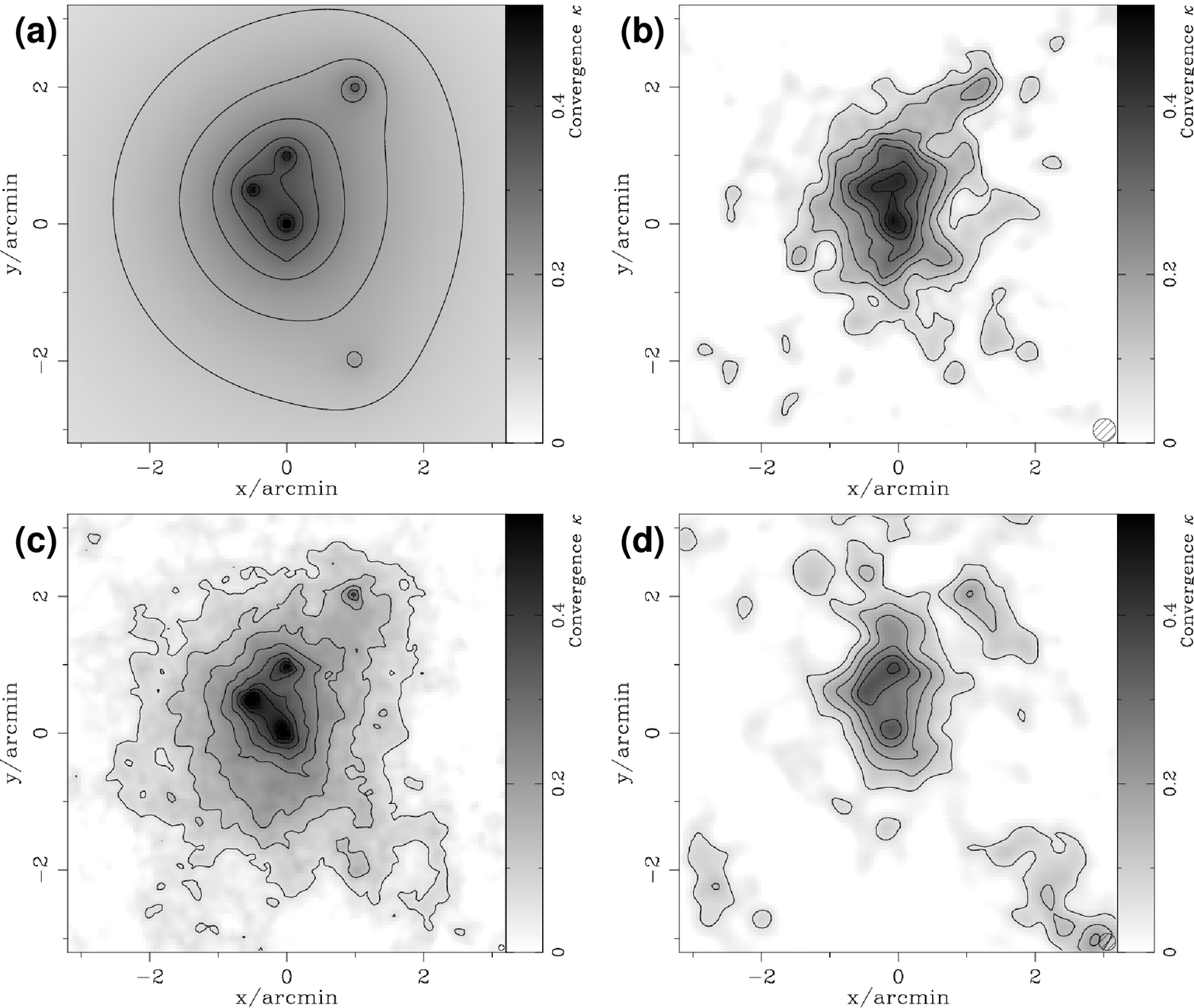}
\label{fig:map}
\caption{
Reconstructions of the dimensionless surface mass density $\kappa$
of a model cluster at $z_d=0.1$
from simulated ellipticity and HOLICs data.
The side length is 6.4 arcmin, and 
$4,096$ ($n_g=100 \, {\rm arcmin}^-2$ )
randomly generated galaxies at a redshift of $z_s=1.2$
have been used for
reconstructions (see \S 4).  
Shown in the top-left panel is the input $\kappa$ map of the model
 cluster. The top-right panel shows a $\kappa$ map reconstructed from
 image ellipticities, $\chi$. The bottom panels show 
$\kappa$ maps reconstructed using HOLICs, 
(c) $\zeta$ and (d) $\delta$.
The lowest contour and the contour interval are $\Delta\kappa=0.06$.
The shaded circle in each panel indicates the FWHM of 
the Gaussian filter.
}
\end{figure}

\begin{figure}
\epsscale{1.0}
\plotone{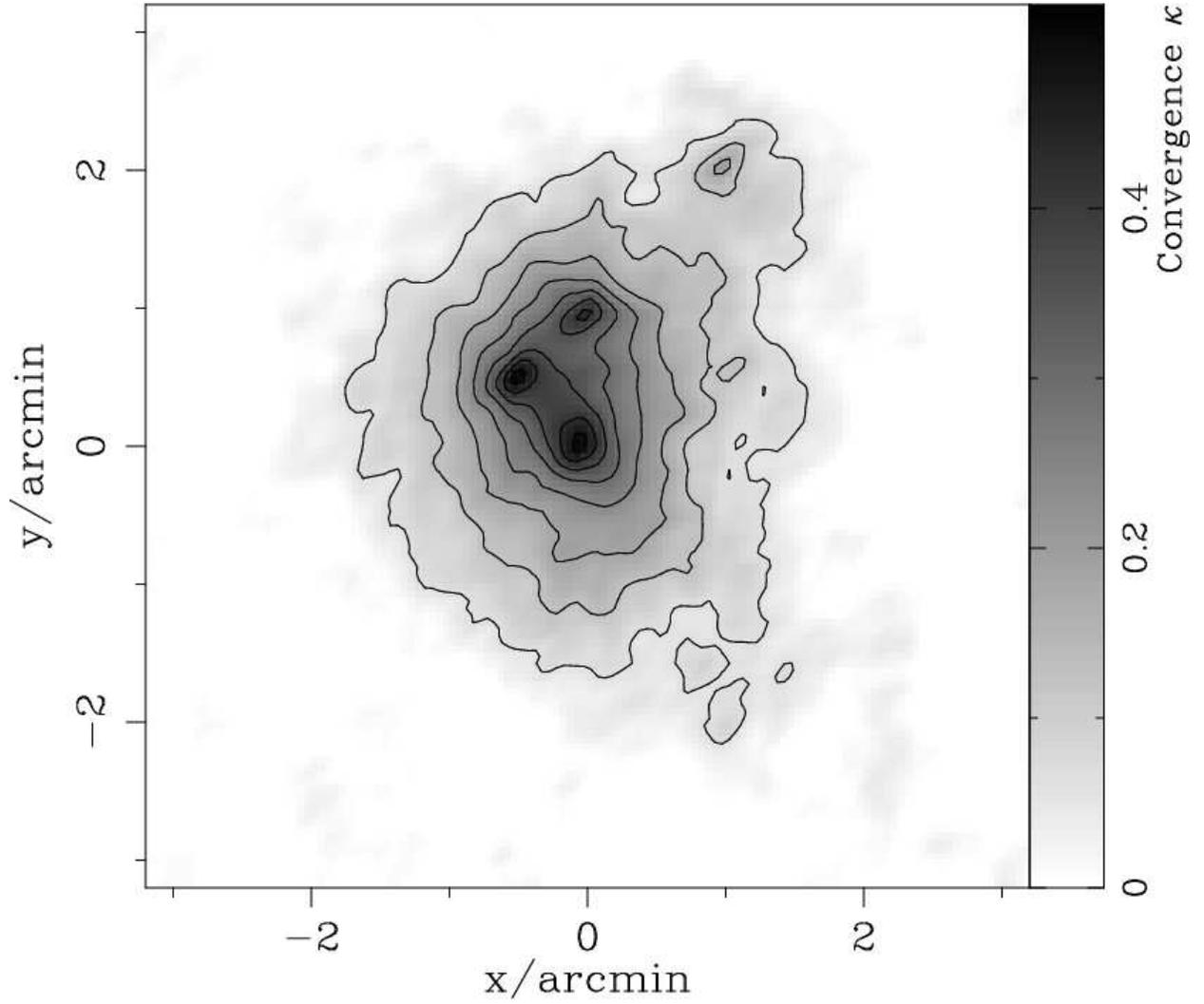}
\label{fig:cmap}
\caption{
Dimensionless surface mass density $\kappa$ 
obtained by combining shear and flexion data in Fourier space (see
 equation [\ref{eq:cmap}]). The lowest contour and the contour interval
 are $\Delta\kappa=0.06$,
}
\end{figure}

\end{document}